\documentclass[onecolumn]{article}

\usepackage{graphicx} 
\usepackage{dcolumn} 
\usepackage{bm} 
\usepackage{wasysym}
\usepackage{textcomp}
\usepackage{stmaryrd}
\usepackage{multirow}
\usepackage{booktabs}
\usepackage[detect-all]{siunitx}
\usepackage{geometry}
\usepackage[sort&compress,numbers]{natbib}
\usepackage{hyperref}
\hypersetup{
	hidelinks}

\begin{document}

\title{Mass spectral analysis and quantification of Secondary Ion Mass Spectrometry data}

\author{A. K. Balamurugan, S. Dash and A. K. Tyagi\\ 
Materials Science Group, Indira Gandhi Centre for Atomic Research \\
Kalpakkam 603102, India.}

\date{}

\maketitle

\begin{abstract}
This work highlights the possibility of improving the quantification aspect of Cs-complex ions in SIMS (Secondary Ion Mass Spectrometry), by combining the intensities of all possible Cs-complexes. Identification of all possible Cs-complexes requires quantitative analysis of mass spectrum from the material of interest. The important steps of this mass spectral analysis include constructing fingerprint mass spectra of the constituent species from the table of isotopic abundances of elements, constructing the system(s) of linear equations to get the intensities of those species, solving them, evaluating the solutions and employing a regularization process when required. These steps are comprehensively described and the results of their application on a SIMS mass spectrum obtained from D9 steel are presented.
It is demonstrated that results from the summation procedure, which covers entire range of sputtered clusters, is superior to results from single Cs-complex per element. 
The result of employing a regularization process in solving a mass spectrum from an SS316LN steel specimen is provided to demonstrate the necessity of regularization.
\end{abstract}

\section{Introduction}
Knowledge of chemical composition is an important aspect in the study of materials. As a surface analytical technique, Secondary Ion Mass Spectrometry (SIMS), is employed for composition profiling of materials in widely varying fields like semiconductor industry\cite{DiffInSiO2} and nuclear technology\cite{Sivai_NaCorr_MMTA2012,David_VoidSwelling_JNM2008}. This technique has high sensitivity, surface specificity, and high dynamic range. However, it lacks quantification because of the dependence of yield of secondary ions on the composition (matrix) of the surface from which it is ejected.\cite{DelineOnMatrixEffect,MingYuOnMatrixEffect} This artefact of the technique is called as matrix effect. In semiconductor industry, the number of commonly analyzed semiconductor materials is limited. Hence, semiconductor research uses matrix-matched standards to quantify SIMS measurements.\cite{LetaIonImplStandards} However, in a general case, like a compound multilayer or an alloy with oxidized surface, the composition is likely to vary over a vast range in the volume analyzed. Such specimens would require a very large number of standards matching each of those compositions to quantify the data. Hence, employing matrix-matched standards in such cases is near to impossible. The matrix effect in the intensities of XCs$_n^+$ secondary ions measured with Cs$^+$ primary ions (where X stands for an atom from the specimen and n is equal to 1 or 2) was shown by Gao \cite{GaoCsM,Gao1Cs2M} to be much lower (in some cases, even by orders of magnitude) than that in secondary elemental ions. However, there is considerable deviation of the composition computed from these XCs$_n$ signals from the actual composition. In spite of developments in understanding the formation process of these species (for example, ref \cite{MageeCW_CsXformation_IJMS1990, Gnaser_XCs_IJMS1998, SmithCsAlInSic, Gnaser_PolarizabilityOnMCs_SSL1994, MootzT_BindEngyOnMCs_IJMS_1992, WitmaackK_AngleOnCsM_NIM_B_1992, Wirtz_OptCsM_CatIonMS_IJMS2003, Kudriavtsev_MCs_mechanism_ASS2003, MineN_CsnM_FromPolymers_ASS_2008}), a gap remains in this approach in reaching complete quantification. The current work provides an incremental step in filling this gap.

\section{Theory}
The XCs$^+$ secondary ions are understood to be formed by the combination of a resputtered Cs$^+$ ion with a sputtered neutral atom from the specimen \cite{MageeCW_CsXformation_IJMS1990,Gnaser_XCs_IJMS1998,SmithCsAlInSic, Gnaser_PolarizabilityOnMCs_SSL1994, MootzT_BindEngyOnMCs_IJMS_1992, WitmaackK_AngleOnCsM_NIM_B_1992, Wirtz_OptCsM_CatIonMS_IJMS2003, Kudriavtsev_MCs_mechanism_ASS2003, MineN_CsnM_FromPolymers_ASS_2008}. Since secondary neutrals are formed as different atomic clusters\cite{MineN_CsnM_FromPolymers_ASS_2008, Goschnick_ClustersFromAerosols, WelzelClusterFormation_TOF_PPM, Belykh_MetalClusterModel}, the intensities of the Cs complexes of all these clusters should be combined to enhance the quantitativeness of XCs SIMS. This was earlier verified with a limited number of Cs complexes \cite{AKB_Comb_XCs_SMT23}. However, testing this with all the Cs complexes involves quantitative measurement of the intensities of the species constituting a mass spectrum and then computing the composition from those intensities.

\section{Material and Methods}
In this report, the details of this process are presented by analyzing a mass spectrum obtained from a sample of D9 steel, produced by M/s. Valinox, France. D9 is the steel used in fast reactors as a construction material of core components because of its resistance to void swelling.\cite{David_VoidSwelling_JNM2008} It was selected as a material for test in this report, because it is a multi-component alloy with known composition. 
The implementation of the above theory involves setting up and solving systems of linear equations to know the composition of the mass spectrum. A mass spectrum is analyzed completely by considering its peaks one after another.
With a peak chosen for analysis, a probable species with its mono-isotopic mass equal to the mass of that peak is first identified. This species could be mono atomic or a poly atomic cluster. Its fingerprint mass spectrum has to be constructed and matched with the experimental mass spectrum. If this species contains $n$ number of elements, the number $N$ of different isotopic combinations forming this species is given by 
\begin{equation}
N=\prod_{i=1}^n\frac{(r_i+s_i-1)!}{r_i!(s_i-1)!}
\label{Eq:Nr of isotopic combination}
\end{equation}
where $r_i$, is the number of atoms of the $i^{th}$ element in the species and $s_i$ is the number of isotopes of the $i^{th}$ element.
Those isotopic combinations with differences in mass, which are not discernible by the resolving power of the mass analyzer, appear as a single peak in the mass spectrum. (For example, in the cluster species CrFe, the combinations, $^{54}$Cr$^{54}$Fe and $^{52}$Cr$^{56}$Fe differ by a mass of 0.003u. A typical mass spectrometer with mass resolving power (MRP) of 500 cannot resolve these two combinations since the MRP required to resolve these species is 35428.) Hence, generally in a mass spectrum, the number of peaks corresponding to a cluster species is fewer than $N$. This group of peaks from the mass spectrometer forms the fingerprint spectrum of the species after normalizing the sum of their intensities to unity. 

The fingerprint spectra of many species are likely to overlap with each other. The difficulty of identifying and measuring the intensities of such species in the measured spectrum depends on the number and complexity of their fingerprint spectra. 
The overlapping fingerprint spectra in the measured spectrum are mathematically represented by a system of linear equations. This system of equations contains one equation for every peak in the spectrum that is the resultant of overlap. 
For $q$ number of species, overlapping with each other, constituting $p$ number of peaks, the system of equations is  
\begin{equation}
\sum_{j=1}^qa_{ij}s_j+\delta_i=m_i
\label{Eq: Equations to be solved}
\end{equation}
where $i$ runs from $1$ to $p$ representing the $p$ number of equations and $m_i$ is the measured intensity of the $i^{th}$ peak. $s_j$ is the intensity of the $j^{th}$ species, to be solved for. If the fingerprint mass spectrum of the $j^{th}$ species has a peak at the mass of the $i^{th}$ peak, $a_{ij}$ is the intensity of that peak of the fingerprint mass spectrum. Otherwise, $a_{ij}$ is zero. $\delta_i$ is the noise that occurred while measuring $m_i$, which is not known. 
This term was added for completeness of mathematical description. However, the above set of equations has to be solved without the knowledge of this term. The above set of equations can be represented in matrix form, after neglecting the noise term as,
\begin{equation}
\mathbf{A\ s=m}
\label{Eq: Matrix form of equations to be solved}
\end{equation}
where $\mathbf{A}$ is the $p \times q$ matrix containing the elements $a_{ij}$ of eqn.\ref{Eq: Equations to be solved}. $\mathbf{s}$ is the $q \times 1$ solution vector and $\mathbf{m}$ is the $p \times 1$ input vector containing $s_j$ and $m_i$ of eqn.\ref{Eq: Equations to be solved} as their elements respectively. The solution providing the least squared deviation between the L.H.S. and R.H.S. of the above equation is obtained by solving the normal form of the above equation,
\begin{equation}
\mathbf{A^T\ A\ s}=\mathbf{A^T\ m}
\label{Eq: Normal form of equations to be solved}
\end{equation}
where $\mathbf{A^T A}$ is a $p \times p$ matrix and $\mathbf{A^T m}$ is a $p \times 1$ vector. $\mathbf{A^T}$ means transpose of $\mathbf{A}$.

In a few circumstances, the solution may turn out to be erroneous (including negative values for the intensities of a few species) due either to the intensity of noise or to wrong choice of species or both. If the error in the solution is due to the noise in the data, the solution can be optimized by following a regularization algorithm such as the iterative algorithm discussed by Gautier \emph{et al}\cite{GautierDecSIA96, GautierDecSIA97} to deconvolve instrumental broadening from SIMS depth profiles. The convolution matrix, solution of deconvolution and measured depth profile found in Ref. \cite{GautierDecSIA96, GautierDecSIA97} are to be replaced by the isotope abundance matrix $\mathbf{A}$, solution for intensities $\mathbf{s}$ and the measured mass spectrum $\mathbf{m}$ respectively to regularize the solution of eqn \ref{Eq: Matrix form of equations to be solved}. Out of the two regularization conditions imposed in Ref. \cite{GautierDecSIA96, GautierDecSIA97}, the condition of positivity has to be retained while rejecting the condition of smoothness since the intensities need not vary smoothly from species to species.
If the choice of species is wrong, the intensities of one or more species in the solution may remain to be negative even after the regularization process. In such cases, other probable species have to be tried out and the process has to be repeated until all the correct species are identified.

The concentration of an element is computed as the fraction of the number of atoms of that element in the Cs complexes to the total number of atoms in all of the Cs complexes excluding atoms of Cs and any other element like O that might originate from the instrument,
\begin{equation}
c_i=\frac{\sum_{j=1}^m n_{ji}s_j}{\sum_{k=1}^n\sum_{l=1}^{e_k} n_{lk}s_k}
\label{Eq: Computing concentration}
\end{equation}
where $c_i$ is the concentration of the $i^{th}$ element, $m$ is the number of species containing the $i^{th}$ element, $n_{ji}$ is the number of atoms of $i^{th}$ element in the $j^{th}$ such species, $s_j$ is the intensity of that ($j^{th}$) species, $n$ is the total number of species identified, $e_k$ is the number of elements (excluding Cs and any other element as mentioned above) in the $k_{th}$ species, $n_{lk}$ is the number of atoms of the $l_{th}$ element in the $k^{th}$ species and $s_k$ is the intensity of the $k_{th}$ species. 

\begin{figure}[h]
\centering
\includegraphics[width=0.8\textwidth]{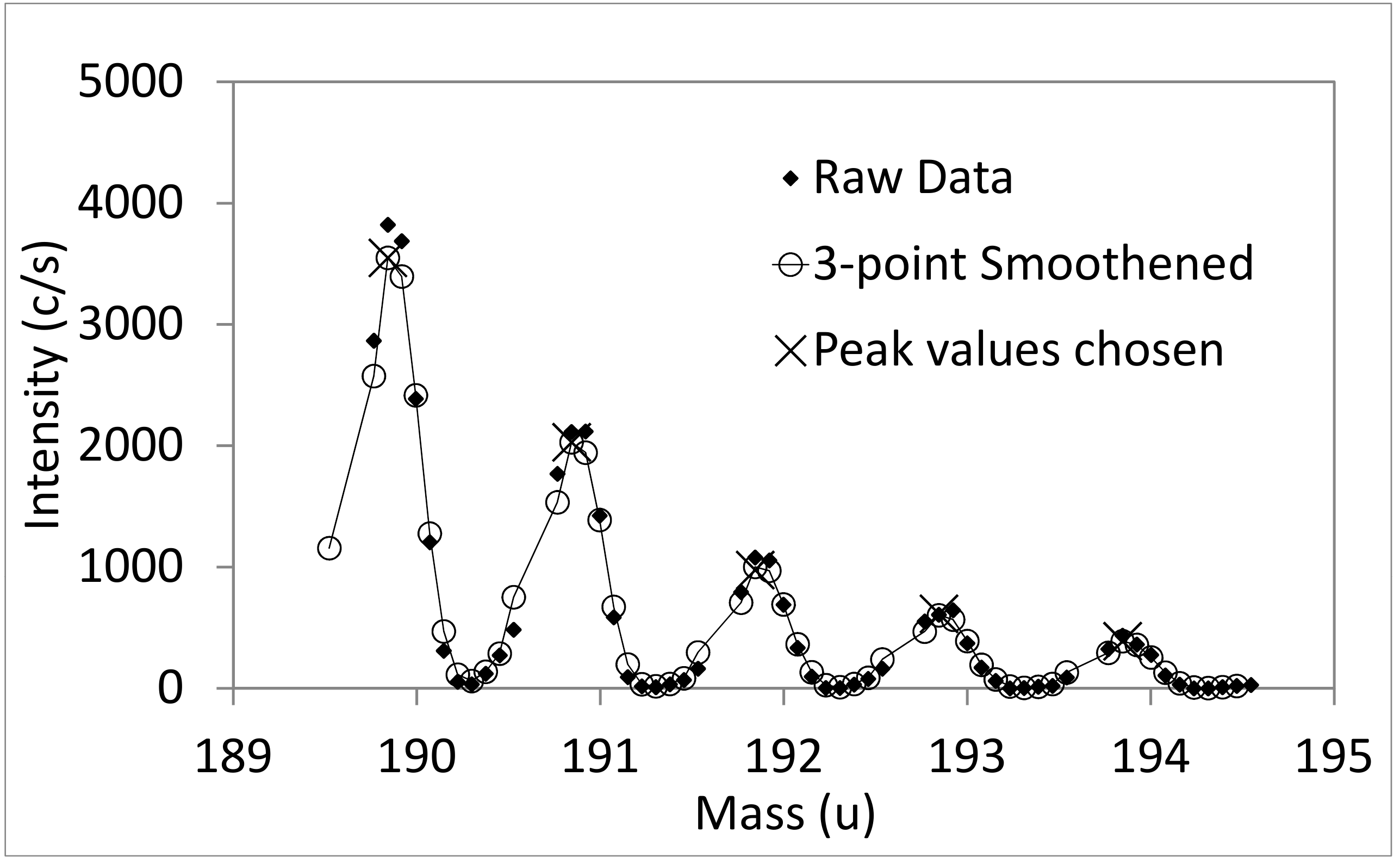}
\caption{\label{Fig: Sample mass spectrum}A portion of raw mass spectrum showing extraction of peak values}
\end{figure}

The process of identifying and measuring the intensities of the species constituting a mass spectrum will be discussed here with a SIMS mass spectrum obtained from a D9 specimen. This mass spectrum was obtained using a SIMS (Cameca ims-4f) system by employing a $20\ nA$, $5.5\ keV$ $Cs^+$ primary ion beam for sputtering the specimen. The primary ion beam was rastered over a square area of width $150\ \mu m$ and the secondary ions were collected from the central circular area of diameter $33\ \mu m$. The mass spectrometer was operated with a mass resolving power of $500$ and energy band-pass of $125\ eV$ for the secondary ions . Eleven data points were recorded around each integral mass number to construct the peaks. A portion of the raw mass spectrum in addition to the peak values as calculated below is shown in figure \ref{Fig: Sample mass spectrum}.

\section{Calculation}
\begin{figure}[h]
\centering
\includegraphics[width=0.8\textwidth]{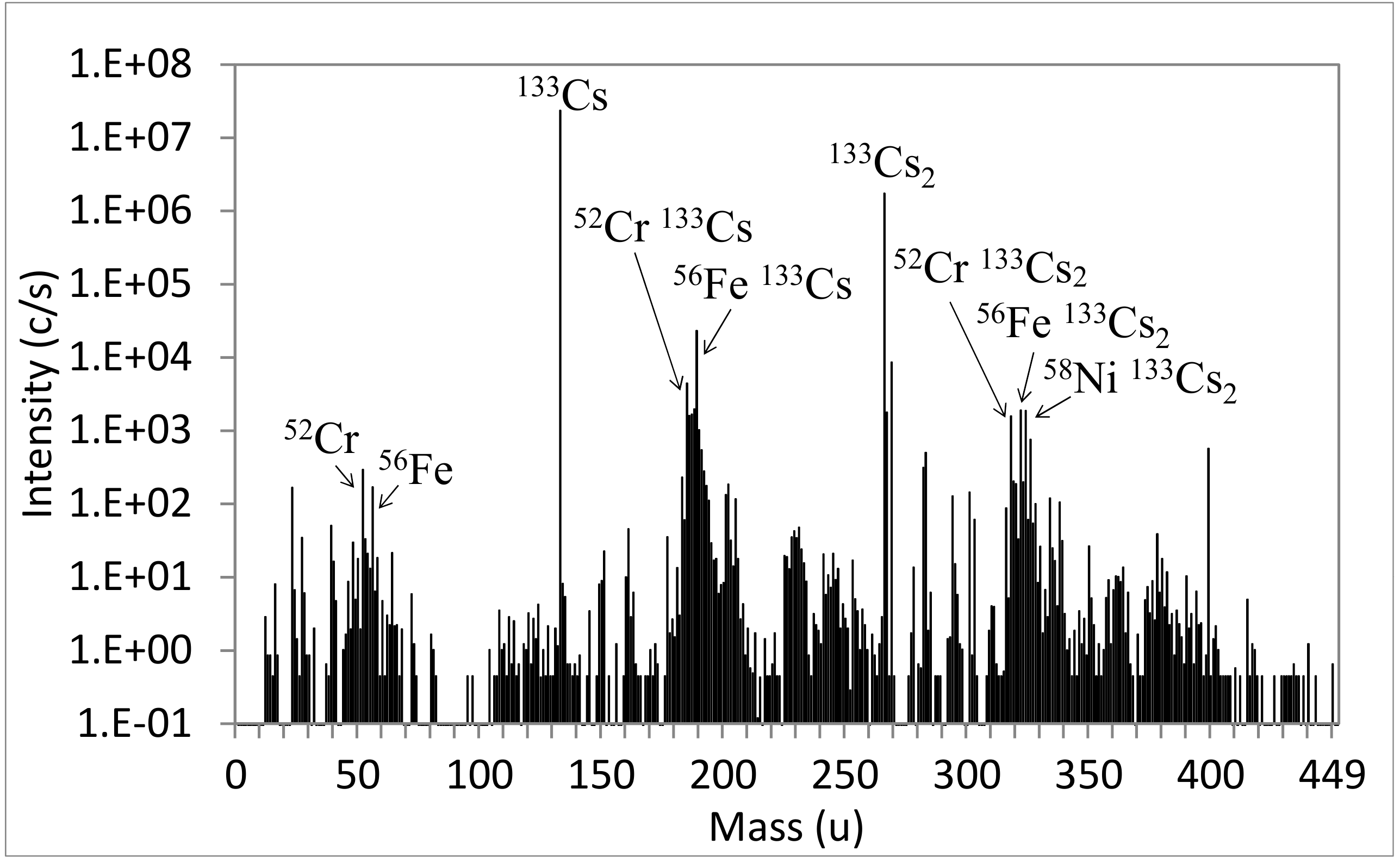}
\caption{\label{Fig: Full mass BarGraph} The mass spectrum over the complete mass range, constructed using the peak values picked up from the raw mass spectrum. The names of a few of the prominent species are labeled to the corresponding peaks.}
\end{figure}

The SIMS mass spectral peaks have the shape of the convolution of the ion-beam crossover with the exit slit of the mass analyzer. The data required for analysis are the heights of these peaks. Since the data is subject to noise and the peaks have a sparse density of data, the highest raw data point of many of the peaks do not represent the apex position. 
Hence, the spectrum was smoothened by 3-point weighted average with the central point receiving a weight of 50\% and the neighboring points a weight of 25\%. After smoothening, the heights of the peaks decreased proportionately and almost all of the peaks obtained a unique apex point as shown in the figure. The apex values of most of the smoothened peaks could be considered as the heights of the respective peaks. The heights of the remaining peaks were estimated manually. The peak values so estimated are shown as a bar graph in figure \ref{Fig: Full mass BarGraph}.

The two strongest peaks in this mass spectrum, at mass numbers 133u and 266u, are those of Cs and Cs$_2$ respectively. The next highest peak is at mass number, 189u that is the mass of $^{56}$Fe$^{133}$Cs. Setting up and solving eqn.\ref{Eq: Matrix form of equations to be solved} for FeCs alone results in a slightly higher estimate for the intensity of FeCs, to minimize the squared deviation from the measured intensity at mass number 187u that is the mass of $^{54}$Fe$^{133}$Cs as well as that of $^{54}$Cr$^{133}$Cs. 
The higher estimate for FeCs compensates the absence of $^{54}$Cr$^{133}$Cs in the equation for mass number 187u. 
After including CrCs in the equation, this error in the estimate for FeCs is corrected. In this manner, the probable species contributing to all the peaks can be tried out one after the other until all the peaks are successfully characterized. 
The final solution is not affected by the order in which the the peaks of the mass spectrum are chosen for analysis. 

\section{Results and Discussion}
\begin{figure}[h]
\centering
\includegraphics[width=0.8\textwidth]{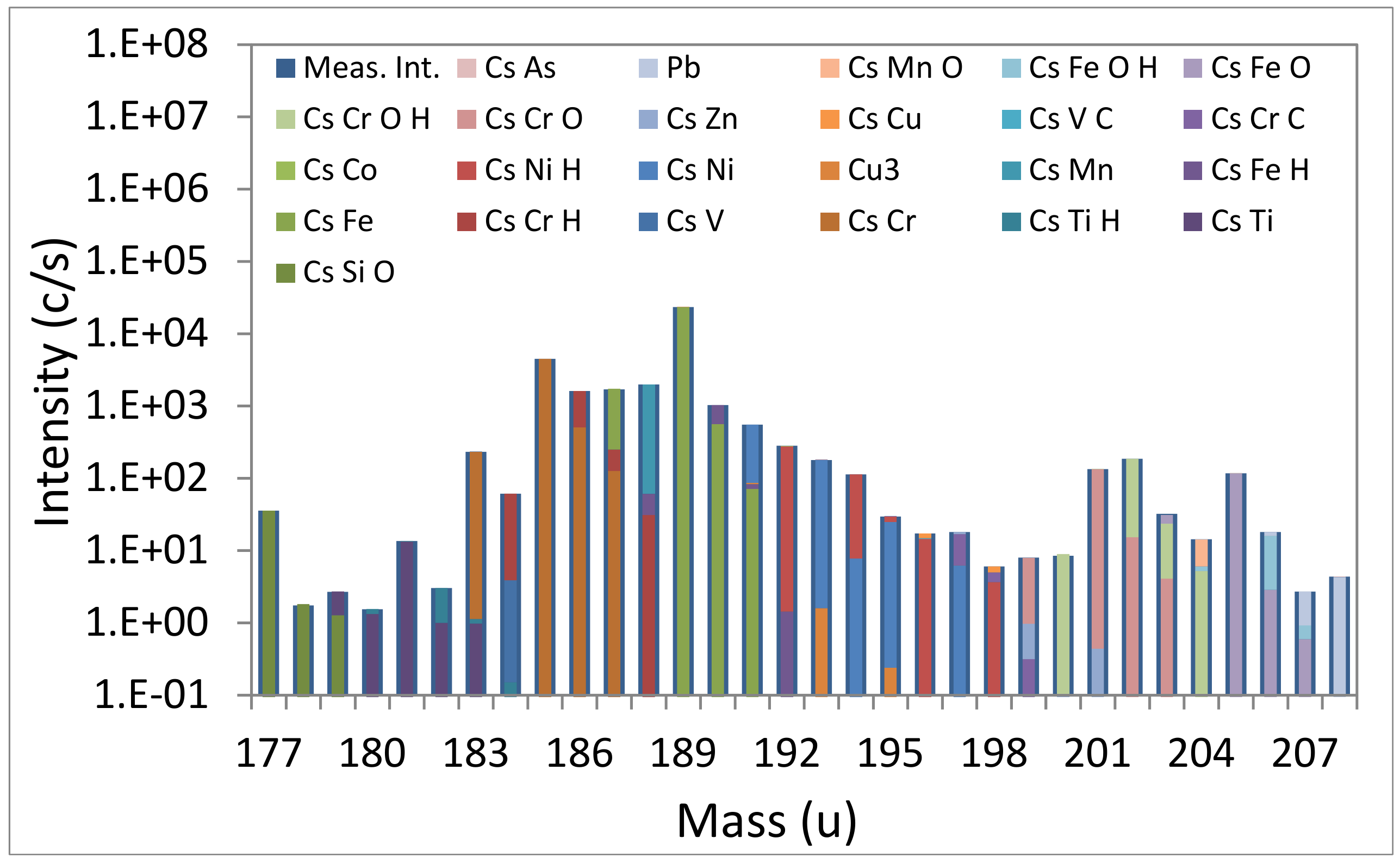}
\caption{A portion of the complete mass spectrum shown in figure \ref{Fig: Full mass BarGraph}, labelled here as ``Meas. MS", and its composition computed using \emph{eqn} \ref{Eq: Normal form of equations to be solved}}
\label{Fig: Solution for species around CsFe}
\end{figure}

In the above mass spectrum, twenty four species were identified overlapping with each other spanning over the mass range from 177u to 208u, as shown in figure \ref{Fig: Solution for species around CsFe}. In this figure, the measured spectrum is shown wider in the background and the constituent fingerprint spectra multiplied by their respective intensities are shown narrower in the foreground. With the perfect solution, the sum of the constituent spectra should be as equal to the measured spectrum as possible as shown in figure \ref{Fig: Solution for species around CsFe}. 
The analysis of the complete mass spectrum resulted in identification of 165 species that are tabulated in Table \ref{table:Species list}. Most of them are Cs complexes that are required for the proposed quantification process.

\newgeometry{margin=1.5cm}

\begin{table} [!htbp]
\begin{center}
\begin{tabular}{|l|r|l|r|l|r|}
	\hline  
	Species 	&	\parbox[m]{1.4cm}{Intensity (c/s)}	&	Species			&	\parbox[m]{0.7cm}{Int. (c/s)}& 	Species			& \parbox[m]{0.7cm}{Int. (c/s)}		\\\hline

	Cs	&	23732613.0	&	Cr O H Cs$_2$	&	13.70	&	Fe Ni O Cs$_2$	&	3.17	\\ \hline
	Cs$_2$	&	1745702.5	&	Fe Cr O Cs$_2$	&	13.47	&	C	&	2.93	\\ \hline
	Fe Cs	&	25458.3	&	Mn	&	13.12	&	Cr Fe H Cs	&	2.86	\\ \hline
	Cr Cs	&	5328.6	&	Cr C Cs	&	12.86	&	Cr O$_2 $H Cs$_2$	&	2.78	\\ \hline
	Ni Cs$_2$	&	2749.8	&	Mo O H Cs	&	10.41	&	Ti H Cs	&	2.77	\\ \hline
	Fe Cs$_2$	&	2094.4	&	Al Cs	&	10.15	&	Ni O Cs$_2$	&	2.77	\\ \hline
	Mn Cs	&	1919.7	&	Fe Cr Cs$_2$	&	9.62	&	Fe Cr N	&	2.72	\\ \hline
	Cr Cs$_2$	&	1892.2	&	Si H Cs$_2$	&	9.58	&	Cu	&	2.64	\\ \hline
	H Cs$_2$	&	1790.9	&	Cr Ni Cs	&	9.42	&	Fe H	&	2.58	\\ \hline
	Cr H Cs	&	1311.1	&	O H Cs	&	9.08	&	Zn Cs	&	2.34	\\ \hline
	Ni Cs	&	685.7	&	O$_3$	&	8.35	&	V O Cs$_2$	&	2.32	\\ \hline
	Cs$_3$	&	574.3	&	H Cs	&	8.16	&	Al$_2$	&	2.19	\\ \hline
	Fe H Cs	&	511.6	&	Mn O Cs	&	8.15	&	Fe$_2 $O Cs	&	2.14	\\ \hline
	O H Cs$_2$	&	507.7	&	O	&	8.14	&	Cr C N Cs$_2$	&	2.01	\\ \hline
	Ni H Cs	&	402.5	&	Cu$_3$	&	8.10	&	W Cs	&	2.00	\\ \hline
	Cr	&	352.8	&	O Cs	&	8.07	&	Fe O$_2 $Cs	&	1.94	\\ \hline
	O Cs$_2$	&	315.6	&	Pb	&	8.05	&	Fe$_2 $O	&	1.82	\\ \hline
	Cr O H Cs	&	204.6	&	Ni C Cs$_2$	&	8.01	&	Cr O$_2 $Cs	&	1.81	\\ \hline
	Cl Cs$_2$	&	197.4	&	Mo H Cs$_2$	&	7.97	&	Mn O Cs$_2$	&	1.80	\\ \hline
	Fe	&	186.2	&	Ni Cr Cs$_2$	&	7.35	&	V	&	1.72	\\ \hline
	Na	&	169.3	&	Si	&	6.60	&	Cu$_2$	&	1.54	\\ \hline
	Fe H Cs$_2$	&	167.4	&	Fe$_2 $O Cs$_2$	&	6.54	&	Al Cs$_2$	&	1.53	\\ \hline
	Cr O Cs	&	160.5	&	Co Cs	&	6.53	&	Cr O N Cs$_2$	&	1.51	\\ \hline
	Mo Cs	&	155.1	&	Na$_2$	&	6.46	&	O$_2$	&	1.39	\\ \hline
	Cr O Cs$_2$	&	144.1	&	Zn	&	6.44	&	Cr$_2$	&	1.37	\\ \hline
	Si Cs$_2$	&	139.6	&	Fe Ni Cs	&	5.72	&	Cr Mn O	&	1.29	\\ \hline
	Fe O Cs	&	128.1	&	Cs H$_2$	&	5.41	&	Ni$_2 $O Cs$_2$	&	1.25	\\ \hline
	Fe O Cs$_2$	&	114.2	&	Cr Fe O Cs	&	5.37	&	Na Cs	&	1.24	\\ \hline
	Mo H Cs	&	107.1	&	Fe Cr O	&	5.27	&	Fe Cu	&	1.24	\\ \hline
	Ni H Cs$_2$	&	74.7	&	Fe Ni H Cs$_2$	&	5.25	&	S Cs$_2$	&	1.10	\\ \hline
	Ti C Cs$_2$	&	63.8	&	F Cs$_2$	&	5.22	&	O H$_2 $Cs$_2$	&	0.99	\\ \hline
	Mo Cs$_2$	&	59.1	&	Co Cs$_2$	&	5.07	&	P Cs$_2$	&	0.94	\\ \hline
	K	&	54.8	&	Fe$_2 $H Cs$_2$	&	5.02	&	O H	&	0.87	\\ \hline
	Si Cs	&	49.9	&	Cr$_2 $Cs	&	5.00	&	N	&	0.87	\\ \hline
	Fe$_2 $Cs$_2$	&	43.2	&	Fe Cr	&	4.70	&	As Cs$_2$	&	0.86	\\ \hline
	Si O Cs	&	38.8	&	Ti H	&	4.61	&	C H	&	0.85	\\ \hline
	Al	&	35.1	&	Ni$_2 $Cs$_2$	&	4.57	&	Fe C	&	0.82	\\ \hline
	Mo O Cs	&	32.4	&	Cr$_2 $O Cs$_2$	&	4.45	&	Co$_2$	&	0.81	\\ \hline
	Cr O$_2 $Cs$_2$	&	32.0	&	Si O Cs$_2$	&	4.42	&	Mn$_2$	&	0.77	\\ \hline
	Fe O H Cs$_2$	&	31.6	&	Cr$_2 $C Cs$_2$	&	4.35	&	Fe$_2 $H	&	0.69	\\ \hline
	Cr H Cs$_2$	&	31.1	&	Fe$_2 $H Cs	&	4.24	&	Si H$_2$	&	0.69	\\ \hline
	Ti	&	28.9	&	Cr O C Cs$_2$	&	4.16	&	Mn O$_2 $Cs	&	0.66	\\ \hline
	Fe Ni Cs$_2$	&	26.5	&	Si O H Cs$_2$	&	4.14	&	P Cs	&	0.66	\\ \hline
	Ni	&	25.4	&	Cr$_2 $O	&	4.08	&	S	&	0.66	\\ \hline
	Cr$_2 $O Cs	&	24.7	&	V Cs$_2$	&	3.89	&	Si H	&	0.61	\\ \hline
	Mn Cs$_2$	&	22.9	&	Ni Cr O Cs$_2$	&	3.86	&	V O$_2 $Cs$_2$	&	0.60	\\ \hline
	H$_2 $O Cs	&	22.8	&	Fe Ni	&	3.75	&	S Cs	&	0.48	\\ \hline
	Cr C	&	22.1	&	V Cs	&	3.73	&	Na Cs$_2$	&	0.45	\\ \hline
	Cr Fe Cs	&	20.3	&	Cu Cr O	&	3.63	&	V$_3$	&	0.41	\\ \hline
	Al$_2 $Cs$_2$	&	20.1	&	C Cs	&	3.52	&	V C Cs	&	0.35	\\ \hline
	Ti Cs	&	18.1	&	V C Cs$_2$	&	3.50	&	Sn Cs	&	0.26	\\ \hline
	Ca	&	17.1	&	Cu Cs$_2$	&	3.35	&	Co	&	0.20	\\ \hline
	Fe O H Cs	&	14.5	&	Cu Cs	&	3.34	&	Ni H	&	0.17	\\ \hline
	Fe$_2 $Cs	&	13.9	&	Fe$_2$	&	3.27	&	As Cs	&	0.06	\\ \hline
	C Cs$_2$	&	13.9	&	Nb Cs	&	3.17	&	Nb Cs$_2$	&	0.06	\\ \hline

\end{tabular}\
\caption{Species identified as constituents of the complete mass spectrum shown in figure \ref{Fig: Full mass BarGraph}}
\label{table:Species list}
\end{center}
\end{table}

\restoregeometry

Composition of D9 computed from Table \ref{table:Species list} using eqn. \ref{Eq: Computing concentration} is shown in Table \ref{table:Composition} in the column named `Multi-species'. The standard composition of D9 is titled as `Known', the composition computed using one Cs complex per element is titled as `Single-species' and the composition computed using atomic secondary ions like Fe$^+$ and Cr$^+$ is titled as `Atomic sec. ions'.
\begin{table} 
\begin{center}
\begin{tabular}{|p{0.8cm}|S[table-format=3.3]S[table-format=3.3]|S[table-format=2.3]|S[table-format=2.3]|S[table-format=2.3]|}
\hline 
\multicolumn{1}{ |l| }{\multirow{2}{*}{\rotatebox[origin=c]{50}{Element }}} & \multicolumn{5}{ c| }{Concentration from} \\ \cline{2-6}
			&	\multicolumn{2}{c|}{Known}	&	 {\parbox[m]{1.5cm}{Multi Cs-species}}	&	 {\parbox[m]{1.5cm}{Single Cs-species}}	& {\parbox[m]{1.5cm}{Atomic sec. ions}}	\\ \hline

	Fe	&	65.22	&					&	64.034	&	71.088	&	27.761	\\ \hline
	Cr	&	16.04	&	{$\pm$\ }0.53	&	20.684	&	14.877	&	52.593	\\ \hline
	Ni	&	13.27	&	{$\pm$\ }0.47	&	8.896	&	7.678	&	3.786	\\ \hline
	Mn	&	1.92	&	{$\pm$\ }0.051	&	4.348	&	5.360	&	1.955	\\ \hline
	Si	&	1.49	&	{$\pm$\ }0.099	&	0.548	&	0.390	&	0.984	\\ \hline
	Mo	&	1.28	&	{$\pm$\ }0.029	&	0.828	&	0.433	&	0.000	\\ \hline
	Ti	&	0.29	&	{$\pm$\ }0.006	&	0.188	&	0.050	&	4.314	\\ \hline
	C	&	0.17	&	{$\pm$\ }0.023	&	0.258	&	0.039	&	0.437	\\ \hline
	Al	&	{\textless}0.10		&		&	0.115	&	0.028	&	5.231	\\ \hline
	V	&	0.049	&	{$\pm$\ }0.003	&	0.031	&	0.010	&	0.257	\\ \hline
	Cu	&	{\textless}0.044	&		&	0.016	&	0.011	&	0.393	\\ \hline
	As	&	{\textless}0.025	&		&	0.002	&	0.000	&	0.000	\\ \hline
	P	&	{\textless}0.025	&		&	0.004	&	0.003	&	0.000	\\ \hline
	Co	&	0.019	&	{$\pm$\ }0.005	&	0.026	&	0.014	&	0.030	\\ \hline
	Pb	&	{\textless}0.019	&		&	0.000	&	0.000	&	1.200	\\ \hline
	W	&	{\textless}0.017	&		&	0.004	&	0.006	&	0.000	\\ \hline
	Nb	&	{\textless}0.010	&		&	0.007	&	0.009	&	0.000	\\ \hline
	S	&	{\textless}0.009	&		&	0.004	&	0.003	&	0.098	\\ \hline
	Sn	&	{\textless}0.003	&		&	0.001	&	0.001	&	0.000	\\ \hline
	Zn	&	{-}					&		&	0.005	&	0.007	&	0.960	\\ \hline

\end{tabular}\
\caption{Comparison of estimates of composition arrived using different methods with the known composition of D9}
\label{table:Composition}
\end{center}
\end{table}
The estimate of composition using single Cs-complex per element shows drastic improvement from the estimate obtained by using atomic secondary ions. This improvement is further enhanced by combining Cs complexes. 
The estimation of the concentration of Fe is very close to the actual concentration and that of Cr exceeds the actual value, while those of most of the other elements show a tendency to approach the actual concentration. 
The formation of Cs complexes is influenced by different parameters. For example, the intensity of an XCs species, where X is an atom of a compound semiconductor or a dopant in Si, is shown to depend on polarizability of X.\cite{Gnaser_PolarizabilityOnMCs_SSL1994} Similar are the results in the case of steel, discussed here. The ratios, intensity of Cs complex to concentration, I$_{XCs}$/C$_X$, of Ni, Cr, Fe and Mn, normalized to the ratio of Fe are,  0.137, 0.852, 1 and 2.52 respectively. 
This is in considerable accordance with Ref \cite{Gnaser_PolarizabilityOnMCs_SSL1994}.
The formation of an XCs$_2$ species is influenced by the electron affinity of X \cite{Gao1Cs2M}. The similar ratios, I$_{XCs_2}$/C$_X$ for the above four elements are respectively, 6.46, 3.67, 1 and 0.371. Thus, the disproportionalities caused by the formation of XCs are countered to some extent by the formation of XCs$_2$. This is overdone in the case of Cr resulting in a net higher estimate for its concentration. In general, combining all the species of the form XCs$_n$, where X is any cluster, is found to take the estimate for composition towards the actual one.

As an example showing the inevitability of the regularization process, the least squared deviation and the regularized solutions for the intensities of a group of species ejected from the surface of an SS316LN sample kept at $550^\circ\mathrm{C}$ for 30,000 hours, is shown in Table \ref{table: Regularization example}. This regularized solution was reached through 27 steps of the iterative regularization process discussed in section 3. This is an example showing how weird the initial solution could be and the capability of the regularization process to arrive at a meaningful solution. However, it should be remembered that in addition to mathematical rigor, physical reasoning should also be followed in the choice of species for analysis.

\begin{table}
\begin{center}
\begin{tabular}{|l|r|r|l|r|r|}
	\hline
\multicolumn{1}{ |l| }{\multirow{2}{*}{Species}} & \multicolumn{2}{ c| }{Solution}	&	\multicolumn{1}{ l| }{\multirow{2}{*}{Species}} & \multicolumn{2}{ c| }{Solution}	\\ 
				\cline{2-3}	\cline{5-6}	
				&	LSD		&	Regul.	&					&	LSD			&	{Regul.}	\\
	\hline

	O Cs$_2$	&	62907	&	62907	&	C$_2 $Cs$_2$	&	-702	&	785	\\ \hline
	Cl Cs$_2$	&	35628	&	35628	&	O$_2 $H Cs$_2$	&	585	&	585	\\ \hline
	Si Cs$_2$	&	32663	&	35397	&	Al Cs$_2$	&	283383	&	480	\\ \hline
	F Cs$_2$	&	20715	&	20715	&	F H$_2 $Cs$_2$	&	417	&	420	\\ \hline
	O$_2 $Cs$_2$	&	20658	&	20664	&	Na Cs$_2$	&	305	&	388	\\ \hline
	O H Cs$_2$	&	20460	&	20460	&	O H$_2 $Cs$_2$	&	339	&	340	\\ \hline
	C$_2 $H Cs$_2$	&	46307724	&	3110	&	F H Cs$_2$	&	222	&	226	\\ \hline
	C$_2 $H$_2 $Cs$_2$	&	-988227	&	2998	&	Cl H Cs$_2$	&	141	&	141	\\ \hline
	P Cs$_2$	&	2139	&	2179	&	Cr$_5 $O$_2$	&	25865	&	38	\\ \hline
	O$_2 $H$_2 $Cs$_2$	&	844	&	844	&	Mn$_2 $O$_3 $Cs	&	-45615716	&	5	\\ \hline
	Fe$_5 $C	&	-5775	&	805	&		&		&		\\ \hline

\end{tabular}

\caption{The least squared deviation (LSD) and regularized (``Regul.")
solutions for a group of species from an SS316LN sample kept at $550^\circ\mathrm{C}$ for 30,000 hours}
\label{table: Regularization example}
\end{center}
\end{table}

\section{Conclusion}
The technique of combining all XCs$_n$ complexes (where X is any cluster) to compute composition is verified to advance the current status of quantification with Cs complexes a step further towards better quantification by two means. One is by the inclusion of atoms in the left out Cs complexes. The other is by the tendency of the disproportionalities of the intensities of XCs and XCs$_2$ species to the concentration of X to counter each other. Delineation of species constituting a mass spectrum, which is a prerequisite for this quantification technique, is aided by the mass spectral analysis described here.

The authors thank Dr. R. Ramaseshan for his help in proof reading and valuable suggestions.


\begin{thebibliography}{99}
\bibitem{DiffInSiO2} Takayuki Aoyama, Hiroko Tashiro, and Kunihiro Suzuki, J. Electrochem. Soc. \textbf{146 (5)}, 1879 (1999).
\bibitem{Sivai_NaCorr_MMTA2012} N. Sivai Bharasi, K. Thyagarajan, H. Shaikh, M. Radhika, A.K. Balamurugan, S. Venugopal, A. Moitra, S. Kalavathy, S. Chandramouli, A.K. Tyagi, R.K. Dayal, and K.K. Rajan, Metallurgical and Materials Transactions A \textbf{43}, 561 (2012).
\bibitem{David_VoidSwelling_JNM2008}C. David, B.K. Panigrahi, S. Balaji, A.K. Balamurugan, K.G.M. Nair, G. Amarendra, C.S. Sundar, Baldev Raj, Journal of Nuclear Materials \textbf{383}, 132 (2008).
\bibitem{DelineOnMatrixEffect}V. R. Deline, William Katz, C. A. Evans Jr., Peter Williams, Appl. Phys. Lett. \textbf{33}, 832 (1978).
\bibitem{MingYuOnMatrixEffect} Ming L. Yu, Wilhad Reuter, J. Vac. Sci. Technol. \textbf{17}, 36 (1980).
\bibitem{LetaIonImplStandards}D. P. Leta , G. H. Morrison, Anal. Chem. \textbf{52 (3)}, 514 (1980).
\bibitem{GaoCsM}Y. Gao, J. Appl. Phys. \textbf{64}, 3760 (1988).
\bibitem{Gao1Cs2M} Y. Gao, Y.Marie, F. Saldi, H.N. Migeon, Proceedings of the 9$^{th}$ International Conference on Secondary Ion Mass Spectrometry - SIMS-IX, Yokohama, Japan, 7-12 November, 406 (1993).
\bibitem{MageeCW_CsXformation_IJMS1990} Charles W. Magee, William L. Harrington, Ephraim M. Botnick, International Journal of Mass Spectrometry and Ion Processes \textbf{103}, 45 (1990).
\bibitem{Gnaser_XCs_IJMS1998} H. Gnaser, International  Journal of Mass Spectrometry  and Ion Processes \textbf{174} 119 (1998)
\bibitem{SmithCsAlInSic}Howard E. Smith, Bang-Hung Tsao, James Scofield, Materials Science Forum \textbf{527-529}, 629 (2006).
\bibitem{Gnaser_PolarizabilityOnMCs_SSL1994} H. Gnaser, H. Oechsner, Surface Science  Letters \textbf{302}, L289 (1994).
\bibitem{MootzT_BindEngyOnMCs_IJMS_1992}T. Mootz, F. Adams, International Journal of Mass Spectrometry and Ion Processes \textbf{152} 209 (1996).
\bibitem{WitmaackK_AngleOnCsM_NIM_B_1992}K. Wittmaack Nuclear Instruments and Methods in Physics Research B \textbf{64}, 621 (1992).
\bibitem{Wirtz_OptCsM_CatIonMS_IJMS2003}T. Wirtz, H.-N. Migeon, H. Scherrer, International Journal of Mass Spectrometry \textbf{225}, 135 (2003).
\bibitem{Kudriavtsev_MCs_mechanism_ASS2003} Yu. Kudriavtsev, A. villegas, A. Godines, R. Asomoza. Appl. Surf. Sci. \textbf{206}, 187 (2003).
\bibitem{MineN_CsnM_FromPolymers_ASS_2008} N. Mine, B. Douhard, L. Houssiau, Applied Surface Science \textbf{255}, 973 (2008).
\bibitem{Goschnick_ClustersFromAerosols} J. Goschnick, M. Fichtner, M. Lipp, J. Schuricht, H.J. Ache, Applied Surface Science \textbf{70/71}, 63 (1993).
\bibitem{WelzelClusterFormation_TOF_PPM} T. Welzel, S. Mandl, K. Ellmer, J. Phys. D: Appl. Phys. \textbf{47}, 065204 (2014).
\bibitem{Belykh_MetalClusterModel} S.F. Belykh, V.I. Matveev, I.V. Veryovkin, A. Adriaens, F. Adams, Nuclear Instruments and Methods in Physics Research B \textbf{155}, 409 (1999).
\bibitem{AKB_Comb_XCs_SMT23} A.K. Balamurugan, S. Rajagopalan, S. Dash, A.K.Tyagi, Proceedings of the 23$^{rd}$ International Conference on Surface Modification Technologies - SMT-XXIII, Mamallapuram, India, November, 537 (2009)
\bibitem{GautierDecSIA96}B. Gautier, R. Prost, G. Prudon, J. C. Dupuy, Surface and Interface Analysis, \textbf{24} 733 (1996).
\bibitem{GautierDecSIA97}B. Gautier, J. C. Dupuy, R. Prost, G. Prudon, Surface and Interface Analysis, \textbf{25}, 464 (1997).

\end{thebibliography}
\end{document}